# Design and modeling of a moderate-resolution astronomic spectrograph with volume-phase holographic gratings

Eduard R. Muslimov*[a], Gennady G. Valyavin [b], Sergey N. Fabrika[b], Nadezhda K. Pavlycheva[a]
[a]Dept. of Optical&electronic systems, Kazan national research technical university named after A.N. Tupolev - KAI, 10 K. Marx str, Tatrstan Rep., Russian Federation 420111; [b]Special astrophysical observatory, Nizhniy Arkhyz, Karachai-Cherkess Rep., Russian Federation 357147

## ABSTRACT

We present an optical design of astronomic spectrograph based on a cascade of volume-phase holographic gratings. The cascade consists of three gratings. Each of them provides moderately high spectral resolution in a narrow range of 83nm. Thus the spectrum image represents three lines covering region 430-680nm. Two versions of the scheme are described: a full-scale one with estimated resolving power of 5300-7900 and a small-sized one intended for creation of a lab prototype, which provides the resolving power of 1500-3000. Diffraction efficiency modeling confirms that the system throughput can reach 75%, while stray light caused by the gratings crosstalk is negligible. We also propose a design of image slicer and focal reducer allowing to couple the instrument with a 6m-telescope. Finally, we present concept of the instrument's optomechanical design.

## 1. INTRODUCTION

Current astronomic spectrographs can be roughly divided into two groups. Instruments from the first group have low or moderate spectral resolution, which corresponds to spectral resolving power ($R=\lambda/\Delta\lambda$) values up to approximately 3000 [1-3]. Such instruments have high throughput of optical system that can reach 82% along all the working spectral range.

The second group unites high-resolution instruments with $R \geq 10000$. It represented mostly by echelle-spectrographs with relatively low throughput, which typically equals to 8% and doesn't exceed the level of 20% even in modern instruments [4].

However, some applications require high optical throughput together with enhanced spectral resolution, which can't be achieved in optical schemes, which are currently in use. Search for very massive stars and black holes of intermediate masses, exoplanets surveys and stellar magnitometry can be examples of such research applications. Thus, in the present paper we consider one of possible concepts of a moderate-resolution spectrograph with high throughput. Our solution is based on a cascade of volume-phase holographic (VPH) gratings. We use such known properties of a VPH grating as high diffraction efficiency (DE) in the working diffraction order, possibility to control the DE curve by adjustment of the holographic layer thickness and modulation depth and to use multiple orders to couple different images or channels [5-7].

The paper is organized as follows. In the section 2 we explain the concept of spectrograph with a VPH grating cascade, in the section 3 we consider an optical scheme of a full-scale spectrograph, designed for 6-m telescope to demonstrate achievable performance. In further sections we focus on a lab prototype of this spectrograph, which is based on a reduced and simplified version of the same optical scheme. We discuss the optical design and image quality (section 4), model the DE of the gratings in cascade and estimate the throughput (section 5). In section 6 the prototype mechanical design is given. Section 7 provides optical design of auxiliary optics for coupling of the instrument with an astronomic telescope. In section 8 we summarize current results and make proposals for future work.

* eduard.r.muslimov@gmail.com; phone +79600381937

## 2. THE INSTRUMENT CONCEPT

In the proposed concept of spectrograph scheme a number of VPH gratings are mounted in a cascade one after another. Each of them works in a certain spectral range. Within this range the grating produces a spectral image in the working +1$^{st}$ order of diffraction. Outside of the range the grating has weak diffraction efficiency, so the rest of radiation is transmitted to the 0$^{th}$ diffraction order. Thus a number of relatively narrow-band spectra covering wide spectral domain with increased dispersion are formed. It's theoretically possible to adjust DE curves of individual gratings in such way, that the working spectral domain will be covered without gaps and with a high throughput for central wavelengths in each sub-range center.

In order to separate the spectra in space a tilt in sagittal plane is introduced on the gratings. It can be implemented in two different ways. Firstly, a grating recorded on a plane-parallel plate can be simply tilted in both of tangential and sagittal planes in order to get required dispersion in each line and separation between them. However, when collimated beam diameter, spectral resolution and camera focal length increase, this approach leads to big inclination angles, too small clearances between gratings and issues with mutual centering of the lines in spectral image. Another option is use of wedges pairs composing plane-parallel plates. The grating is imposed to the inner (glued) surface of such assembly. So when accounting for the 0$^{th}$ order of diffraction this component will work as a usual plate, while in the working +1$^{st}$ order it will produce spectrum with calculated angles of deviation. This second approach definitely brings more technological difficulties, but it allows to align a few moderate-resolution spectra on the same detector array. Thus we will use it for a full-scale instrument scheme, while the first approach is applicable for the reduced prototype.

In general, the optical scheme design algorithm consists of the following steps. The initial data are: full working spectral range, overall scheme length, detector array size and desired reciprocal linear dispersion. On the first stage the sub-ranges for each grating are chosen and collimator and camera focal lengths are defined. Then the gratings' grooves frequencies and their tilt angles in two planes are approximately calculated from the required dispersion and image lines spacing. On the next stage the collimator and camera optical schemes are developed. Finally the entire optical scheme is modeled and optimized using a standard numerical method. During this optimization all the conditions and limitations such as maintenance of linear dispersion, spectral image lines spacing and centering, as well as the image quality are accounted for. Let us note here, that we allow a certain turn of the image lines to present here. It should be taken into consideration during further processing of the spectra.

## 3. FULL-SCALE OPTICAL SYSTEM

### 3.1 Optical scheme

We consider a spectrograph optical scheme for the visible domain 430-680 nm. It's divided to three sub-ranges: 430-513, 513-597 and 597-680 nm. We assume that the detector array has sensitive area of 36x24 mm (or more) with pixel size equal to 5.2x5.2 µm. Besides, the length of each spectrum line should be 30 mm, i.e. the reciprocal linear dispersion is 2.77 nm/mm. The collimator and camera focal lengths are 170 mm each. Finally, the equivalent F/# is 3.7 that should be enough to cover exit aperture of the most of existing astronomical telescopes and provide moderate resolution and high image.

Using the listed initial data and applying the described procedure we obtain the following results. The gratings parameters are given in the Table.1

Table 1. Parameters of VPH gratings for the full-scale spectrograph.

| Sub-region | Grooves frequency | Tangential tilt | Sagittal tilt |
|---|---|---|---|
| 430-513 nm | 1616 mm$^{-1}$ | -10,007° | 13,951° |
| 513-597 nm | 1337 mm$^{-1}$ | 1,258° | 9,116° |
| 597-680 nm | 1049 mm$^{-1}$ | 15,053° | 14,612° |

The collimator lens is a usual triplet, while the camera lens is triplet-based system with an additional field corrector. It also should be noted that the lines spacing in the image is more than 2.2.mm.

The scheme general view is presented on Fig.1. The system overall dimensions are 755x440x170 mm.

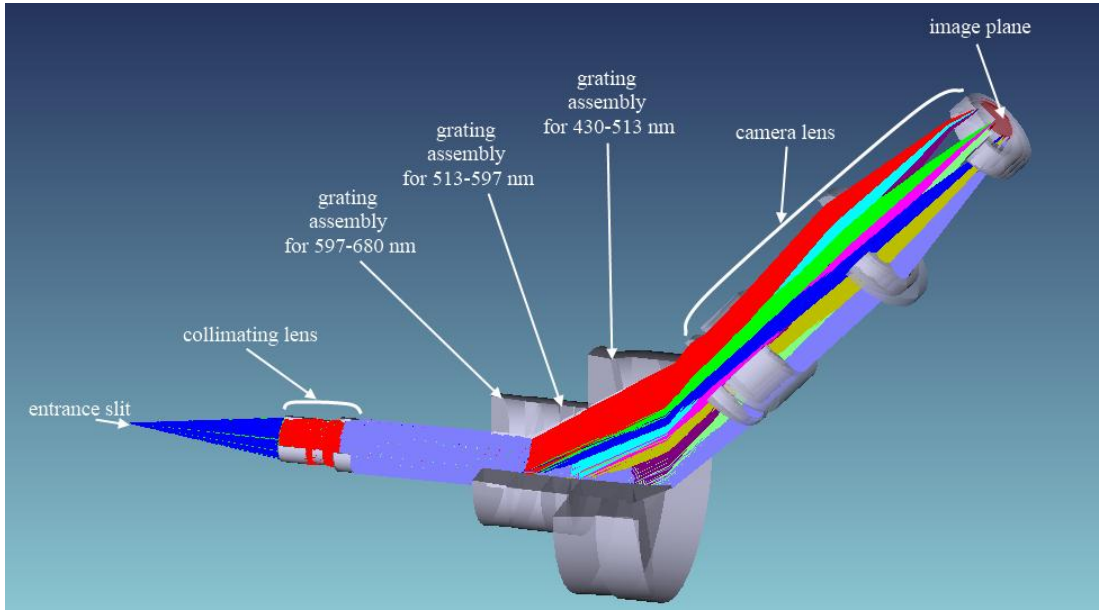

Figure 1. General view of the full-scale spectrograph scheme.

### 3.2 Image quality and spectral resolution

To demonstrate the image quality achieved in the scheme its' spot diagrams are presented on Fig. 2. It's clear from the diagrams that a good aberration correction is provided for the entire working domain. Modeling shows that if the entrance slit width is 30 μm, the instrument function's FWHM is equal to 31-56 μm; 31-45 μm and 30-34 μm for each of the sub-ranges starting from the short-wave one. It means that the spectral resolution is 0.082-0.147 nm; 0.083-0.098 nm and 0.086-0.124, respectively. In other words, the spectral resolving power in relative units for the spectrograph scheme is 5243-7906. It agrees with the definition of a moderate resolution spectrograph, which was given above.

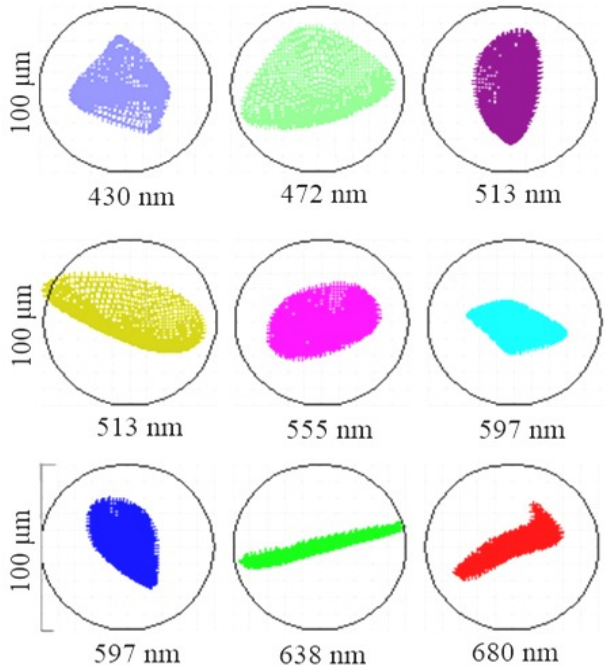

Figure 2. Spot diagrams for the full-scale spectrograph scheme.

Thus the developed optical scheme provides moderately high spectral resolution and has a relatively compact size and small number of refracting surfaces. However, an experimental testing is strongly needed for proof of the concepts and assumptions designed-in the scheme. The described scheme is not appropriate for building of a testing prototype because of complexity and high cost of the high aperture wedge based dispersive assemblies. So a simplified and reduced in size version of the spectrograph with VPH grating cascade was developed.

## 4. OPTICAL SYSTEM OF THE SPECTROGRAPH PROTOTYPE

### 4.1 Optical scheme

The simplified optical scheme for the spectrograph prototype is based on three VPH grating working in the same sub-ranges as that in the full-scale scheme. Each of the grating is imposed on a plane-parallel plate made of BK7 glass and protected by a cover glass (the substrate thickness is 2.6 mm and the cover glass thickness is 2.2 mm). The target length of spectral image was set to 20 mm (so the reciprocal linear dispersion is 4.15 nm/mm). The required lines spacing is the image was between 1.5 and 4 mm. We propose to use two identical commercial Tessar-type lenses as the collimator and camera. In this case we use lenses with focal length of 135 mm and F/# of 2.8. The entrance aperture was assumed to be decreased to F/# = 4.

After the design procedure identical to the described above we obtain the following gratings parameters (see Table 2).

Table 2. Parameters of VPH gratings for the spectrograph reduced prototype.

| Sub-region | Grooves frequency | Tangential tilt | Sagittal tilt |
|---|---|---|---|
| 430-513 nm | 1726 mm$^{-1}$ | 28,333° | 3,387° |
| 513-597 nm | 1523 mm$^{-1}$ | 22,851° | 1,409° |
| 597-680 nm | 1205 mm$^{-1}$ | 1,701° | 3,885° |

It should be emphasized here that in such scheme a special control should be provided for the gratings tilt and gaps between them. Additional boundary conditions were introduced to the merit function during the optimization to ensure incidence angles appropriate for the DE maximization and realizable scheme geometry.

The scheme general view is presented on Fig.3. The overall dimensions are 313x178x54 mm.

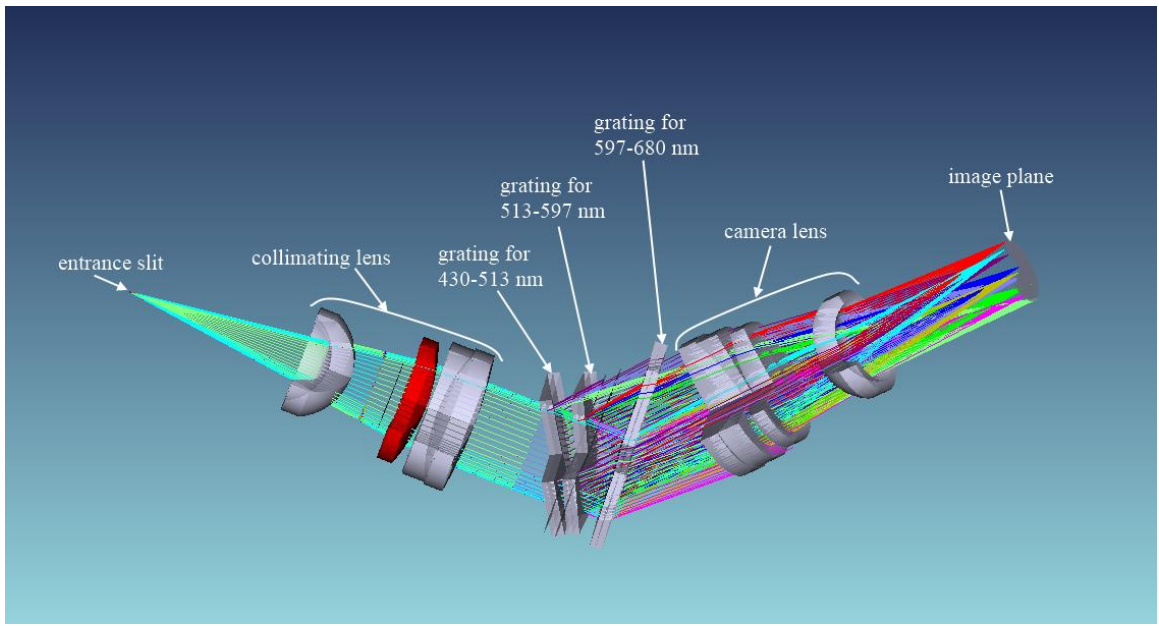

Figure 3. General view of the spectrograph reduced prototype.

Note, that in this case the gratings are assembled into cascade in an inverse order in comparison with the first scheme. in addition a certain vignetting can be seen from the figure. It doesn't exceed 2,5% and can't affect the spectrograph performance.

### 4.2 Image quality and spectral resolution

Similarly to the previous case we provide spot diagrams to demonstrate the achieved image quality (Fig.4). Except of some residual aberrations on the field of view edges the chosen lenses provides relatively high image quality.

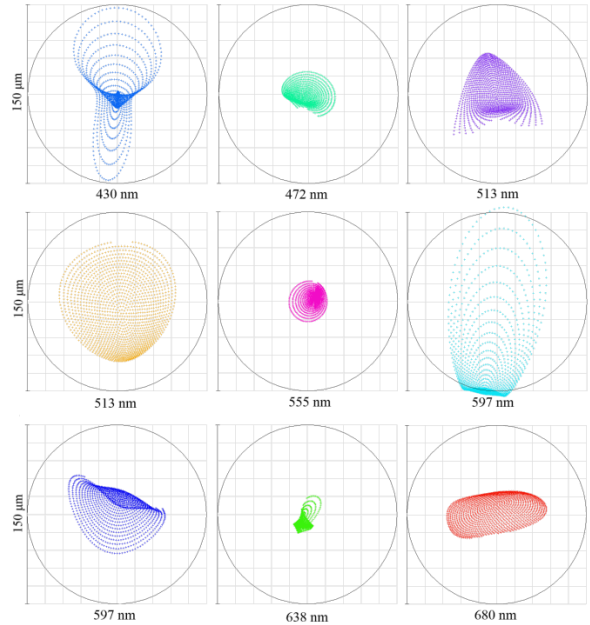

Figure 4. Spot diagrams for the spectrograph reduced prototype.

For the entrance slit width of 30 μm the following instrument functions FWHM values were obtained: 30-49 μm; 30-79μm and 30-36 μm from the short-wave sub-range to the long-wave one. The corresponding spectral resolution is 0.125-0.203 nm; 0.125-0.330 nm and 0.125-0.151, respectively, while the spectral resolving power is 1553-5124. Thus, the reduced optical scheme provides image quality high enough for demonstration and proof of the operation principle.

## 5. THROUGHPUT ESTIMATION

For estimation and optimization of the gratings diffraction efficiency Kogelnic coupled wave theory is used [8]. An approach, similar to that used in [9] is applied. It's assumed that the holograms are recorded on BCG layers. After the holographic layers parameters optimization we obtain that for technologically achievable values of the layer thickness of 10-30 μm and modulation depth of 0.01-0.025 very high diffraction efficiency and spectral selectivity can be reached.

The obtained DE curves are presented on Fig.5. After subtracting of the gratings crosstalk, surface reflection losses and so on, we derive that maximum throughput in the gratings cascade can reach approximately 75% when on the sub-ranges borders it will count about 40%.

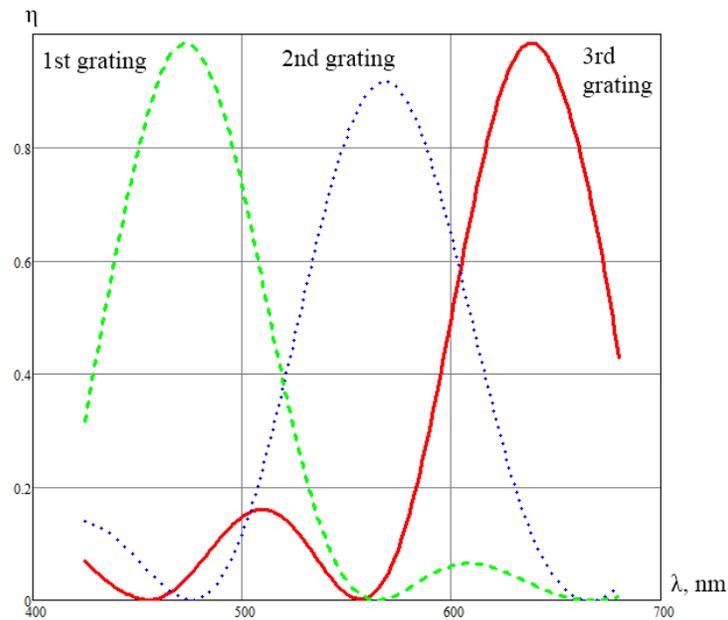

Figure 5. Diffraction efficiency curves for the VPH gratings in cascade.

## 6. OPTOMECHANICAL DESIGN OF THE PROTOTYPE

On the basement of the developed optical scheme an optomechanical design of the spectrograph prototype was made. It consists of a customized grating assembly, two commercial lenses, entrance slit and camera with their holders. The gratings assembly is customized, while the rest of the parts can be composed of commercial elements. The grating assembly consists of a common base with three stairs determining the gratings tilt angles in the tangential plane, plates, which are used for positioning in the sagittal plane, and frames holding each of the gratings. Plates are attached to the base with spherical-head bearing screws. Spring disks and screws with spherical points are used for precise alignment of the gratings angular positions. Two different views of the assembly are shown on Fig.6 (the slit, camera and lens holders are removed).

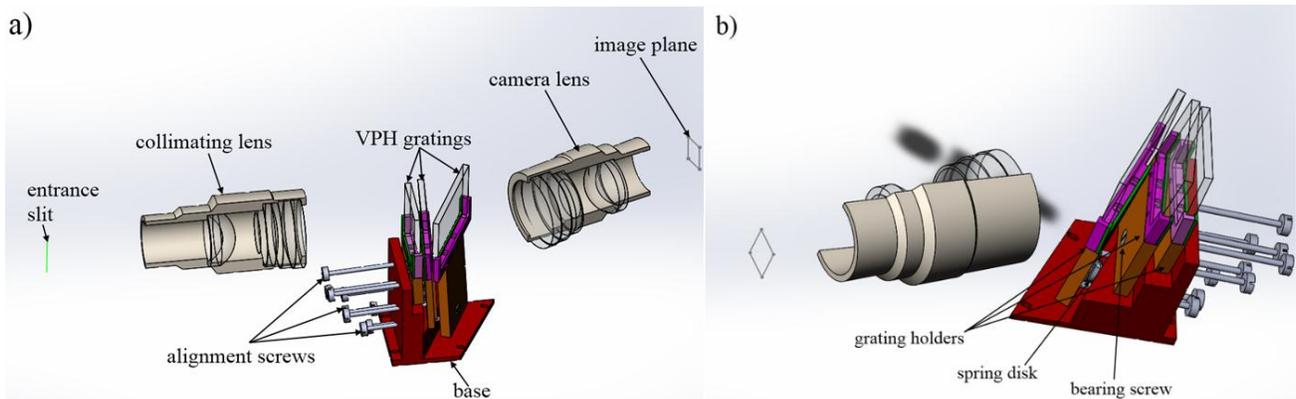

Figure 6. Solid model of the optomechanical design of the spectrograph prototype: a) general view; b) detailed view of the alignment mechanism.

Such a classical design is notable because for its' manufacturability and flexibility in use. Currently, all the necessary parts drawings are prepared and the assembly is ready for production.

## 7. COUPLING OPTICS

Real performance of the developed spectrographs will depend on coupling with the previous optical system. Thus, actual illuminance will depend on matching between the telescope exit aperture and the spectrograph entrance aperture. On the other hand, the spectral resolution will be determined by the entrance slit width, which should be in agreement with the telescope image. Specialized projection lenses or focal reducers used to solve the first task and image slicers are used for the latter one.

On the current stage of our project we consider mounting of the spectrograph in lateral Nasmyth focus of a 6-m F/4 telescope as the most realistic option. In this case the exit beam has equivalent F/# of 31, and the image diameter can reach 1.5 mm [10].

Accounting for the spectrograph parameters we found that the focal reducer should have linear magnification equal to -0,13$^x$. It should be corrected for chromatism in all the working domain 430-680 nm. The geometric image diameter will be about 195 μm, so the slicer should divide the image at least to 7 stripes.

### 7.1 Focal reducer

As soon as the entrance aperture of the focal reducer under consideration is quite small the optical scheme can be relatively simple. We consider an all-refractive scheme consisting of two components with a parallel beam between them. The first component represents an uncemented doublet and the second one is triplet lens. For simplicity reasons we assume that the doublet focal length is 310 mm, so the triplet has focal length of 40 mm. We optimize the optical system for 20 mm-wide field of view.

The image quality analysis shows that the RMS radius of spot diagrams is 3.5-3.6 μm and energy concentration within a 30 μm-circle is 96%.

### 7.2 Image slicer

In our case the image slicer can be mounted in a diverging beam in front of the first lens of reducer. It would allow us to make the slicer design as simple as possible and refuse of any optical power on its' elements. Also, strict requirements for the optics throughput should be kept in mind.

Considering the listed features a slicer scheme based on total internal reflection (TIR) was chosen [11, 12]. It represents an cemented assembly of a prism and plate. One of the prism facets is inclined in such a way, that the TIR condition is violated in a certain zone and the incident beam is divided to a few stripes.

In the developed slicer the plate and prism are made of BK7 glass. The plate thickness is 2 mm, prism cathetus length is 18 mm, distance between the Nasmyth focus and the entrance facet is 10 mm and the side facet tilt is 7.93°.

Operation of the coupling optics in whole was modeled with ray-tracing software. It was assumed that all the glass surfaces have anti-reflection coatings with residual reflection of 2%. Fresnel losses and multiple reflections were accounted for. In addition volume absorption with attenuation index of 0.0017 cm$^{-1}$ was introduced to the model. The slicer operation is shown on Fig.7a, while on Fig. 7b general view of the coupling optics is given.

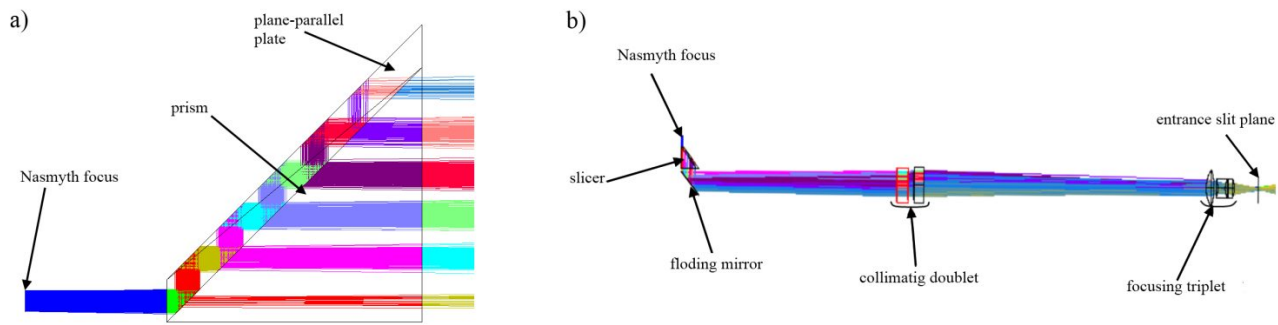

Figure 7. Optical system for coupling of the spectrograph with a telescope: a) operation of the image slicer; b) general view.

The image after coupling optics is shown on Fig.8. The overall efficiency is 69,4%, that is relatively high for such systems and appropriate for first tests.

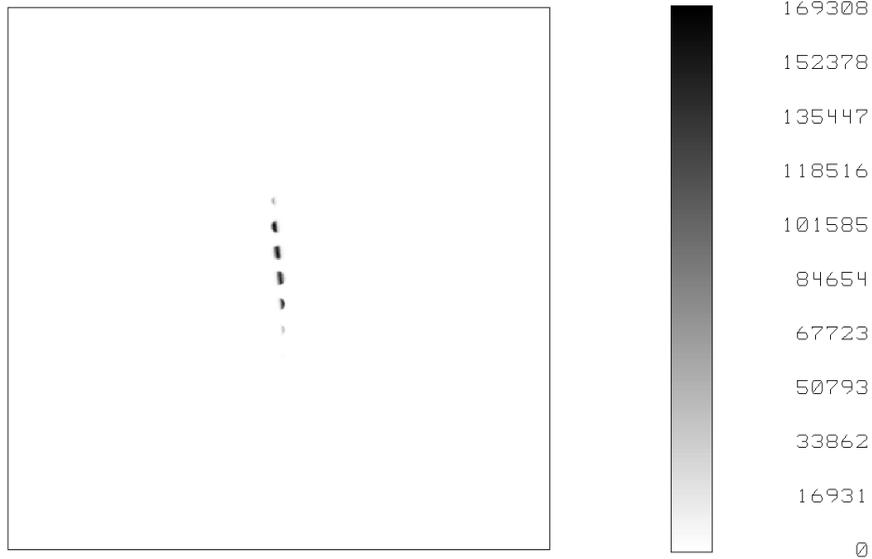

Figure 8. Image after the focal reducer and image slicer.

## 8. CONCLUSIONS

In the present work we proposed a concept of new spectral instrument for astrophysics. It's based on a cascade of volume-phase holographic gratings with aligned diffraction efficiency curves. Each of the gratings serves as a dispersive and splitting element thus allowing to form a spectral image consisting of a few lines and provide moderately high spectral resolution along with high throughput. Spectrographs of such a kind could be very effective in spectral observations of exoplanets (for example [13]), magnetometric observations [14], details of brightness distribution on surfaces of faint degenerate stars ([15]), etc.

We discussed two possible implementations of this concept - a full-scale scheme with gratings imposed on wedges and a reduced scheme with gratings on plane substrates. For each of them we carried out computer simulation and demonstrated the achieved spectral resolution. For the second one we presented results of diffraction efficiency estimation and optomechanical design. Finally, an optical schemes of focal reducer and image slicer necessary to couple the spectrograph with 6-m telescope Nasmyth focus were developed.

The next step of the research will be production and testing of the spectrograph reduced prototype in order to proof its' basic operation principles and demonstrate performance.

## ACKNOWLEDGEMENTS

This work was supported by the Russian Science Foundation (project No. 14-50-00043).

## REFERENCES

[1] I. Appenzeller, K. Fricke, W. Fuertig et al., "Successful Commissioning of FORS1 - the First Optical Instrument on the VLT," The Messenger, 94, 1 (1998)

[2] T. Szeifert, I. Appenzeller, W. Fuertig et al. , "Testing FORS: the first focal reducer for the ESO VLT.," Proc. of SPIE, 3355, 20 (1998)

[3] V.L. Afanasiev, & A.V. Moiseev, "The universal focal reducer of the 6-m telescope SCORPIO," Astronomy Lett., 31, 194 (2005)

[4] Kukushkin D. E., Sazonenko D. A., et al., "High-resolution fibre-fed spectrograph for the 6-m telescope. Polarimetric unit," Astrophysical Bulletin , 71 (2), 249-256 (2016).

[5] Barden, S. C., Arns, J. A., Colburn, W. S., & Williams, J. B., "Volume-Phase Holographic Gratings and the Efficiency of Three Simple Volume-Phase Holographic Gratings," Publications of the Astronomical Society of the Pacific, 112(772), 809–820, (2000).

[6] Blanche P, Gailly P, Habraken S, Lemaire P, Jamar C, "Volume phase holographic gratings: large size and high diffraction efficiency,". Opt. Eng.,43(11),2603-2612 (2003).

[7] Palmer, C. and Loewen, E. [Diffraction grating handbook], Newport Corp, Rochester, 170-130 (2005)..

[8] Kogelnic, H. "Coupled wave analysis for thick hologram gratings," Bell Syst. Tech. J., 48, 2909-2947 (1969)

[9] Muslimov, E.R. "Transmission holographic grating with improved diffraction efficiency for a flat-field spectrograph,"   Proc. of SPIE, 8787, 87870B (2012)

[10] D. N. Monin, V. E. Panchuk , "A moderate-resolution Nasmyth-focus spectrograph of the 6-m BTA telescope," Astronomy Letters, 28 (12), 847-852 (2002).

[11] Gerardo Avila, Carlos Guirao, Thomas Baader, "High efficiency inexpensive 2-slices image slicers,". Proc. SPIE ,8446 , 84469M (2012).

[12] Akito Tajitsu, Wako Aoki, and Tomoyasu Yamamuro, "Image Slicer for the Subaru Telescope High Dispersion Spectrograph," Publ Astron Soc Jpn, 64 (4),77(2012)

[13] G. G. Valyavin , A. O. Grauzhanina et al., "Search for signatures of reflected light from the exoplanet HD 189733b by the method of residual dynamical spectra", Astrophysical Bulletin, 70(4), 466-473 (2015)

[14] I. I. Romanyuk, "Magnetic fields of chemically peculiar and related stars. Main results of 2014 and near-future prospects," Astrophysical Bulletin, 70  (2), 191-205 (2015).

[15] G. Valyavin,    D. Shulyak,    G. A. Wade et al. , "Suppression of cooling by strong magnetic fields in white dwarf stars," Nature, 515, 88–91 (2014)